\documentclass[pra,12pt]{revtex4}

\usepackage{amsmath}    
\usepackage{graphicx}   

\begin{document}

\title{Absorption spectrum in the wings of the potassium second
resonance doublet broadened by helium}
\author{Fran\c{c}ois Shindo, James F. Babb, Kate Kirby, and Kouichi Yoshino}
\affiliation{Harvard-Smithsonian Center for Astrophysics,\\
60 Garden St., Cambridge, MA 02138}

\begin{abstract}
  We have measured the reduced absorption coefficients occurring in
  the wings of the potassium $4\,^2\!S$--$5\,^2\!P$ doublet lines at
  404.414~nm and at 404.720~nm broadened by helium gas at pressures of
  several hundred Torr.  At the experimental temperature of 900~K, we
  have detected a shoulder-like broadening feature on the blue wing of
  the doublet which is relatively flat between 401.8~nm and 402.8~nm
  and which drops off rapidly for shorter wavelengths, corresponding to
  absorption from the $\textrm{X}{}^2\Sigma^+$ state to the
  $\textrm{C}{}^2\Sigma^+$ state of the K--He quasimolecule.  The
  accurate measurements of the line profiles in the present work will
  sharply constrain future calculations of potential energy surfaces
  and transition dipole moments correlating to the asymptotes
  $\textrm{He--K}(5p)$, $\textrm{He--K}(5s)$, and $\textrm{He--K}(3d)$.
 
\end{abstract}

\maketitle

The broadening of atomic spectral lines through collisions with
ambient atoms is of wide-ranging importance and
study of the absorption line
shape profiles can reveal much about the absorbers' interactions with
the perturbing atoms and with their physical environments.
Spectroscopy of the resonance lines of alkali-metal atoms pressure
broadened by various elements (including like-species) at temperatures
up to 1000~K has revealed that the far wing absorption profiles possess
shapes such as satellite peaks and shoulders indicative of the
interactions between absorber and perturber, see, for example,
Refs.~\cite{ChePhe73,LorNie77,VezMovPic80,SzuBay96,ShuParYos00,GreHamCro06}.  
Most studies of 
far wing broadening have centered on the principal series
resonance line for an alkali-metal atom in the presence of a ground
state inert gas atom.  Broadening of the second 
(and higher) principal lines  in
these cases has been less frequently studied.

For potassium, there are evidently few experimental studies of the
absorption coefficients in the far wings of the second principal
doublet $4\,^2\!S$--$5\,^2\!P$ in the presence of helium.  Delhoumme
\textit{et al.} measured the absorption coefficients in the presence
of neon at 533~K~\cite{delhoume_quasi-static_1981} in an experiment
closely related to the theoretical calculations by
Masnou-Seeuws~\cite{masnou-seeuws_model_1982} of K--He and of K--Ne
potential energy curves and transition dipole moments for states
correlating to K$(4s)$ and K$(5p)$.  Dubourg and
Sayer~\cite{dubourg_experimental_1986} focused their absorption
measurements on the red wing in the presence of helium at 548~K.
Earlier work using spectrographic methods identified
the wavelength at which there is a break off in 
the satellite band intensity on the blue side of the second
doublet~\cite{CheWil61,JefWil65}.  Cantor, Penkin and
Shabanova~\cite{CanPenSha85} measured the absorption coefficients of
the second doublet in the presence of helium at temperatures of around
570--655~K.  We report in this letter spectroscopic measurements of
the far blue and red wings of the potassium doublet line
$4\,^2\!S$--$5\,^2\!P$ broadened by helium at 900~K.

\begin{figure}[ht]
        \begin{center}
                \includegraphics[bb = 86 38 450 728,  width=.49\textwidth, angle=-90]{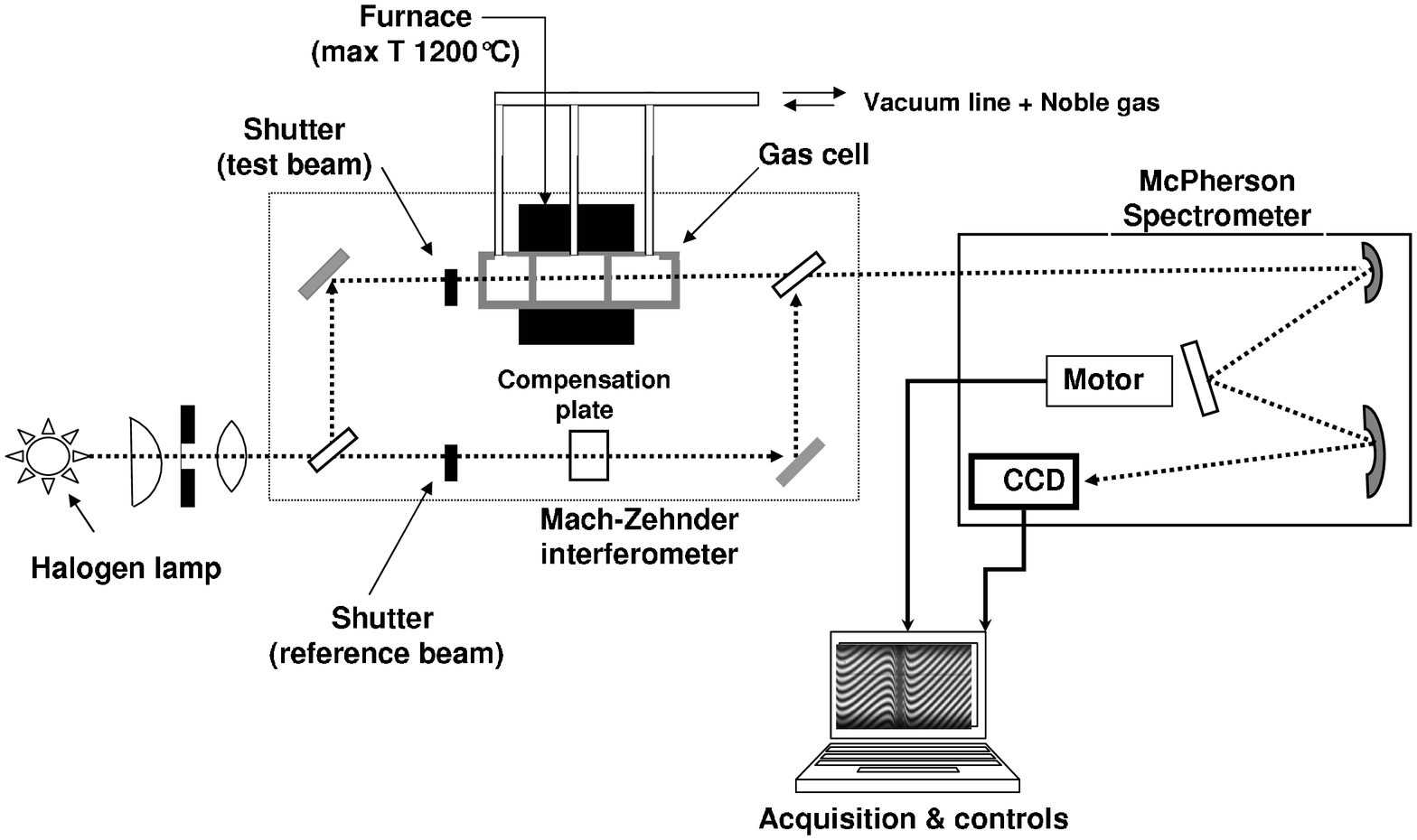} 
        \end{center}
        \caption{Experimental set-up as described in the text.}
        \label{fig:setup}
\end{figure}

The critical elements in the absorption spectroscopy experiment,
illustrated in Figure~\ref{fig:setup}, are a Mach-Zehnder
interferometer, a 3~m Czerny-Turner grating spectrograph (McPherson
model 2163) equipped with a CCD camera detector (Andor 1024$\times$256 pixels
with a size of 26 $\mu$m), and a gas cell designed for the study of
hot and corrosive vapors. This gas cell allows measurements at
different buffer gas pressures from 10 Torr to 1000 Torr and
temperatures up to 900~K~\cite{Shindo-unpub}. The optical path length
of the cell is delimited by two MgO windows separated by a distance of
20$\pm$0.01~cm. A uniform temperature along this path length is
maintained by a split tube furnace.  For a continuum light source, we
use a tungsten halogen lamp (250~W, 24~V) with the beam collimated
through an arrangement of optics before entering the interferometer.
The gas cell is included in one of the arms of the interferometer. In
the other arm, we have placed a stack of windows identical to those
contained in the gas cell to compensate for the delay introduced by
the optics of the cell. The interferometer is adjusted at optical zero
path difference to produce a set of horizontal interference fringes
which are focused by a lens on the entrance slit of the spectrometer.
Each arm of the interferometer can be blocked by a shutter. Thus, we
can measure the spectrum emerging from the gas cell and a reference
spectrum emerging from the other arm which serves as a calibration to
correct effects due to the variation of the light source intensity
with time. The association of the interferometer and the spectrometer
is essential to apply the so-called ``hook'' or anomalous dispersion
method which provides a measurement of the atomic number density of
the absorbers vaporized in the gas cell.

The hook method uses the change in the refractivity in the vicinity of
an atomic spectral line to determine the quantity $Nlf$, where $N$ is
the atomic number density, $l$ is the optical path length and $f$ is
the oscillator strength of the transition producing the spectral line.
If $l$ and $f$ are known, then the density $N$ can be determined
from the measured value of $Nlf$.
This method and its applications are extensively described by
Parkinson~\cite{parkinson_laboratory_1987},
Marlow~\cite{marlow_hakenmethode_1967}, and Huber and
Sandeman~\cite{huber_measurement_1986}.  
As discussed previously, the interferometer is adjusted at
optical zero path difference with the gas cell inserted in one arm.
In this configuration, we can observe through the spectrometer a set
of near horizontal interferences fringes which are spatially
distributed along the spectrometer slit.  If a dispersive medium
exhibiting a resonance line at a certain wavelength $\lambda_{0}$ is
introduced into the gas cell, then the variation of the refractive
index with wavelength around $\lambda_{0}$ causes a distortion of the
fringe profile near the resonance line.  By inserting an additional
compensation plate into one of the interferometer arms, the fringes
are shifted to higher orders, and the deformations appear as hooks on
each side of the resonance line.  The equations describing the shape
of the fringes in the image plane of the spectrometer and extensions of
the formalism to the case of hooks formed under the influence of more
than one absorption line are given in
Refs.~\cite{parkinson_laboratory_1987} and
\cite{huber_measurement_1986}.
The vertical position $y_k (\lambda)$ of a given fringe of order $k$
(along the spectrometer slit) is a function of the wavelength
$\lambda$. For a resonance line doublet, it can be expressed
with four free parameters $\alpha$, $\beta$, $\delta$ and $\zeta$ as:
\begin{equation}
\label{eq:fringe2}
y_{k} (\lambda) ={\alpha}+{\beta}{\lambda}+{\delta}k{\lambda}+{\zeta}\left(\frac{A_1}{\lambda_{1}-\lambda}+\frac{A_2}{\lambda_{2}-\lambda}\right),
\quad\text{with}\quad A_{i}=\frac{r_{0}N_\textrm{K}f_{i}l{\lambda}_{i}}{4{\pi}}. 
\end{equation} 
In the configuration of the spectrometer, only the fringes
forming on each side of the first resonance doublet around 770~nm offer a
sufficient dispersion and visibility to apply the hook method for the
determination of $N_\textrm{K}$,
so in our case $\lambda_1=766.491$~nm and $\lambda_2=769.898$~nm.
 To retrieve the atomic number density $N_\textrm{K}$ of potassium atoms
vaporized in the gas cell, we apply a least-squares fit to the 
shapes of the hooked fringes 
recorded on the CCD camera, using
equation~\eqref{eq:fringe2}.  The atomic potassium density $N_\textrm{K}$ is
then simply calculated from the ratio \textit{$\zeta$/$\delta$}. The
wavelength and oscillator strength of each line are taken from
Hasegawa \textit{et al.}~\cite{hasegawa_relation_1991}.

For this experiment, the spectrometer is equipped with a 1200 line/mm
plane ruled grating working in first order. A stepper controller
can rotate the grating to 116 different fixed angles
enabling the spectrometer to cover the wavelength range from 360~nm 
to 
920~nm. For each angle of the grating position, a spectrum 
covering about 50~nm can be recorded
by the CCD camera. 
Each such spectrum was calibrated against 
available lines from a selection of different standard hollow
cathode lamps (Na--Ar, Ti--Ne, Ca--Ne, and Co--Ne). Backlash of the motor
driving the rotation of the grating causes an uncertainty of $\pm$~5
pixels in the localization of the calibration lines. In the region of
the 404~nm doublet, this uncertainty corresponds to an error in the
wavelength of about 0.04~nm.  As an independent check, we measured the
forbidden lines $4\,{}^{2}S_{1/2}$--$3\,{}^{2}D_{3/2}$ and
$4\,{}^{2}S_{1/2}$--$3\,{}^{2}D_{5/2}$ \textit{after} calibration and
found each differed from the previously
reported~\cite{striganov_tables_1968} positions of 464.188~nm and
464.237~nm, respectively, by only 0.004~nm.  

We calculate the absorption coefficient $\tau_{\lambda}$ using Beer's law: 
\begin{equation}
\label{eq:beerlaw}
\tau_{\lambda}=-\frac{1}{l}
\ln\left(\frac{S^\textrm{K}_{\lambda}}{R^\textrm{K}_{\lambda}}{\times}\frac{R^\textrm{E}_{\lambda}}{S^\textrm{E}_{\lambda}}\right). 
\end{equation}
In equation~\eqref{eq:beerlaw}, the quantity $S^\textrm{K}_{\lambda}$ refers to the
intensity emerging from the gas cell in presence of potassium vapor and helium. Prior to the vaporization of the
potassium in the gas cell, we record a set of backgrounds denoted as
$S^\textrm{E}_{\lambda}$ when the gas cell is empty. The ratio between the spectra $R^\textrm{K}_{\lambda}$ and $R^\textrm{E}_{\lambda}$ account for the variation of the light source measured in the reference beam.
Assuming that only helium and potassium are involved in the absorption
spectrum, in equation~\eqref{eq:beerlaw} we can separate
${\tau}_{\lambda}$ into two linear terms involving atomic densities $N_\textrm{K}$ and $N_\textrm{He}$, and
reduced absorption coefficients: 
\begin{equation}
\label{eq:redcoeff}
\tau_{\lambda}=N_\textrm{K}N_\textrm{He}{\gamma}_\textrm{K--He}+N^{2}_\textrm{K}{\gamma}_\textrm{K--K}. 
\end{equation}
The first term describes the
absorption resulting from the collisions between helium and potassium
atoms, $\gamma_\textrm{K--He}$, and the second term describes the
pure potassium absorption, $\gamma_\textrm{K--K}$. The atomic number
density of helium atoms is calculated from our measurement of the gas
pressure according to the following expression derived from the gas
law:
 \begin{equation}
\label{eq:gaslaw}
N_\textrm{He}=9.657{\times}10^{18}
  \frac{P_\mathrm{Torr}}{T_\mathrm{Kelvin}}\mathrm{cm}^{-3}. 
\end{equation}
The reduced absorption coefficients can be therefore determined by measuring ${\tau}_{\lambda}$ at several pressures of helium and/or different densities of potassium.

The stability of the potassium atomic number density in the gas cell
is checked before and after the collection of each
spectrum. When equilibrium is reached in the gas
chamber, the pressure fluctuation of helium lies in a range of 1\% or
less around the initial amount, and we observe the same stability for
the temperature of the gas cell. For these conditions, the atomic
number density of potassium varies within a range of 5\%. For a cell
temperature of 900~K, we collected several spectra at pressures of
helium in the range of 200--900~Torr, and with $N_\textrm{K}$ in the range of
$10^{15}$--$10^{16}$ cm$^{-3}$.  We have noticed that the continuum
varies from one spectral measurement to another indicating an
irregular decrease of the transmission even when only helium gas is
present in the cell. We have established that this phenomenon is
related to the gradient of pressure between the gas chamber and the
rest of the gas cell even in the absence of potassium vapor in the
gas cell. It may be caused by a decrease in the load rate of the
springs due to the temperature or the thermal expansion of the
windows. In order to obtain absolute absorption coefficients, we have
assumed a baseline for our spectra using the experimental spectrum and
calculations of Dubourg and Sayer~\cite{dubourg_experimental_1986}.
Their measurements of absorption coefficients shows that the
wavelength domain 420--430~nm is free of K--K and K--He absorption.
Therefore, we can use this wavelength interval in our spectrum to
establish our experimental baseline.

\begin{figure}[b]
        \begin{center}
                \includegraphics[bb = 86 38 578 728,  width=.49\textwidth, angle=-90]{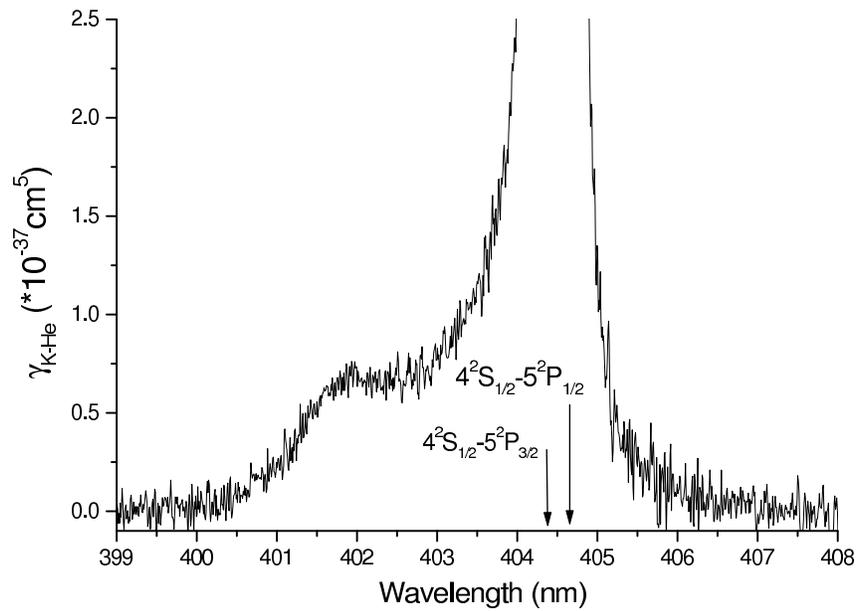} 
        \end{center}
        \caption{Reduced absorption coefficient $\gamma_\textrm{K--He}$  at 900~K.  For values less than $1{\times}10^{-37}\,\textrm{cm}^5$ 
the uncertainty is 15 percent and for greater values saturation makes it difficult to assign a consistent uncertainty.}
        \label{fig:404nm}
\end{figure}

For the ${\gamma}_\textrm{K-He}$ coefficient profile measured at 900~K
(Fig.~\ref{fig:404nm}), 
we observe a shoulder-like feature 
on the blue wing of the doublet line with an almost
flat intensity of
$6.8{\times}10^{-38}$ cm$^{5}$ between 401.8 and 402.8~nm that drops
off rapidly for shorter wavelengths. This profile results from an average of
several measurements with a standard deviation of 15\%.  The
calculations of Masnou-Seeuws~\cite{masnou-seeuws_model_1982} for the
K--He and the K--Ne potentials lead to the prediction of classical
satellites on the blue wing of the second resonance doublet. In the
case of the K--Ne system both theory and experiment exhibit
 agreement to within 1~cm$^{-1}$ in the observation of a shoulder-like satellite
feature at 403.2~nm\cite{delhoume_quasi-static_1981}.  For the K--He
system, from
the potential energy curves calculated  by Masnou-Seuuws, the classical
satellite occurs at around 401.5~nm which differs by 19~cm$^{-1}$ from
our result. Earlier studies, which used photometric analysis of spectrograms, located the peak intensity of the satellite band at 401.76(3)~nm~\cite{CheWil61} and at 401.69(6)~nm~\cite{JefWil65}, but the peak wavelength determination using this method can be affected by the possible presence of background molecular potassium bands. At a
wavelength of 402.5~nm, Johnson and Eden~\cite{johnson_continua_1985}
measured a signature probably due to $\textrm{K}_2$ in the absorption
spectrum of a pure potassium sample. But in our case, considering our
experimental densities of potassium and helium, the K--He contribution
is 500 to 1000 times stronger than the expected contribution from
$\textrm{K}_2$. In the present experiment the potassium dimer
contribution is ten times weaker than the noise level of 0.4~\% in our
transmission spectra.  Therefore, we can reasonably associate our
shoulder-like feature to quasi-molecular
absorption from the
$\textrm{X}{}^2\Sigma^+$ state to the
$\textrm{C}{}^2\Sigma^+$ state of the K--He system.
After the completion of the
present analysis, we discovered a study published in the Russian
literature reporting a similar experiment on the broadening of the K
4s--5p doublet by helium and other inert gases in the temperature range
of 570--655~K~\cite{CanPenSha85}. These authors  identify a
satellite feature in absorption at 401.85(5)~nm 
in good agreement with our observations 
at 900~K.

The collisions between potassium and helium give rise to a broadening 
feature on the blue side of the $4\,^2\!S$--$5\,^2\!P$ second
resonance doublet line at around 401.8~nm,
in agreement with prior studies~\cite{CheWil61,JefWil65,CanPenSha85}.
This is consistent with the prediction from 
previously calculated potential energy
curves~\cite{masnou-seeuws_model_1982}
of  a classical satellite feature 
around this wavelength.
Further theoretical calculations of the
potential energy curves and transition dipole moments
for the molecular states correlating to 
the asymptotes He--K$(5p)$, He--K$(5s)$, and He--K$(3d)$
would enable more precise predictions of the absorption coefficients.
This work is the first result in an extensive study of the
absorption arising from K--He and K--H$_{2}$ gas mixtures. A paper reporting our results on the broadening of the 770~nm doublet and
their application of this work to astrophysical models of the atmospheres of brown dwarfs and certain class of extrasolar planets is currently 
in preparation~\cite{Shindo-unpub}.

This work was supported in part by NASA under grants NAG5-12751 and
NNG06GF06G issued through the Science Mission Directorate, Universe
Division, Astronomy and Physics Research and Analysis program.  
We thank 
Dr. L. Gardner,  Dr. D. Fabricant,
 and in particular Dr. W. H. Parkinson for valuable advice.


\section*{}

\end{document}